\RequirePackage{fix-cm}
\documentclass[twocolumn]{svjour3}          
\smartqed  
\usepackage{graphicx}
\journalname{Journal of Radioanalytical Nuclear Chemistry}
\begin{document}

\title{Excitation functions of $^{nat}$Pb(d,x)$^{206,205,204,203,202}$Bi, $^{203cum,202m,201cum}$Pb and $^{202cum,201cum}$Tl reactions up to 50 MeV
}


\author{F. Ditr\'oi \and F. T\'ark\'anyi \and S. Tak\'acs \and A. Hermanne \and  A.V. Ignatyuk
}


\institute{F. Ditr\'oi F. T\'ark\'anyi \and S. Tak\'acs \at
              Institute for Nuclear Research, Hungarian Academy of Sciences \\
              Tel.: +36-52-509251\\
              Fax: +36-52-416181\\
              \email{ditroi@atomki.hu}           
           \and
           A. Hermanne \at
              Cyclotron Laboratory, Vrije Universiteit Brussel (VUB), Brussels, Belgium
           \and
              A.V. Ignatyuk \at
            Institute of Physics and Power Engineering (IPPE), Obninsk 249020, Russia  
}

\date{Received: 2014 / Accepted: 2014}

\maketitle

\begin{abstract}
Cross-sections of deuteron induced nuclear reactions on lead were measured up to 50 MeV using the standard stacked foil irradiation technique and high resolution $\gamma$-ray spectrometry. Experimental cross-sections and derived integral yields are presented for the $^{nat}$Pb(d,x)$^{206,205,204,203,202}$Bi, $^{203cum,202m,201cum}$Pb and $^{202cum,201cum}$Tl reactions. The experimental data were compared with the results from literature and with the data in the TENDL-2013 library (obtained with TALYS code). The cross-section data were analyzed also with the theoretical results calculated by using the ALICE-IPPPE-D and EMPIRE-D codes.

\end{abstract}


\section{Introduction}
\label{intro}
In the frame of our systematic study of activation cross-sections of deuteron induced nuclear reactions for different applications and for development of the reaction models \cite{1,2} we have investigated the activation cross-sections on lead. In an earlier study we have already presented the excitation functions for longer lived activation products up to 40 MeV \cite{3} ($^{nat}$Pb(d,xn)$^{203,204,205,206,207}$Bi, $^{nat}$Pb(d,x)$^{203}$Pb, $^{202}$Tl reactions). In this work we have extended the energy range up to 50 MeV and we have tried to obtain data also for shorter-lived products.  A literature search resulted in only one additional cross-section study reported by Wasilyeski et al.  \cite{4}  up to 27 MeV while experimental integral yield data in  the 11- 22 MeV  energy range were measured by Dmitriev et al. \cite{5}.

\section{Experimental}
\label{sec:2}
The experimental techniques and data analysis were similar or identical as described by us in recent publications (c.f.  \cite{6,7,8,9}). Here we present shortly the most important facts, and details specific for this experiment.
Taking into account the large energy range to be covered the foil stacks were irradiated at two incident energies (50 MeV and 21 MeV). The first stack (series 1), containing $^{nat}$Eu (100 $\mu$m), Sb(2 $\mu$m) evaporated onto Kapton foil (13 $\mu$m), Pb (15.74 $\mu$m ) and Tb (95.47 $\mu$m) target foils interleaved with 49.06 $\mu$m Al monitor foils was irradiated at the Cyclone 90 cyclotron of the Université Catholique in Louvain la Neuve (LLN) with a 50 MeV incident energy deuteron beam (60 min, 150 nA). A second stack (series 2) containing  $^{nat}$Pb (15.74 $\mu$m) target foils, Al energy degrader/monitor foils (50 $\mu$m), was irradiated at the CGR 560 cyclotron of the Vrije Universiteit Brussel (VUB) with a 21 MeV incident energy deuteron beam (60 min, 105 nA).
The activity produced in the targets and monitor foils was measured non-destructively (without chemical separation) using a high resolution HPGe $\gamma$-ray spectrometer. For the high energy irradiation four series of measurements were performed starting at about 4 h, 50 h, 450 h and 2200 h after EOB at 50, 25 and 5 cm source-detector distances. 
Also four series of $\gamma$-spectra measurements were done for the low energy experiment starting at about 1 h, 4 h, 45 h and 770 h after EOB and at 20, 15 and 5cm source-detector distances. 
The decay data of the investigated activation products, taken from the online database NuDat2 \cite{10} are summarized in Table 1 together with the possibly contributing reactions and their thresholds. Effective incident beam energy and the energy scale were determined primary by calculation \cite{11} and finally by using the excitation functions of the $^{24}$Al(d,x)$^{24}$Na  and $^{nat}$Ti(d,x)$^{48}$V monitor reactions \cite{12} simultaneously re-measured over the whole energy range.  The beam intensity (the number of the incident particles) was initially obtained through measuring the charge collected in a short Faraday cup and, if needed, adapted on the basis of the monitor reactions.
The curve of the re-measured excitation function in comparison with the IAEA recommended data \cite{8} can be found in Fig. 1.	
The uncertainty of the median energy in each foil was estimated taking into account the cumulative effects of possible uncertainties (primary energy, target thickness, energy straggling, correction to monitor reaction). The uncertainty of the cross-sections was obtained as the sum in quadrature of all linear contributions of systematic uncertainties (beam current (7\%), beam-loss corrections (max. of 1.5\%), target thickness (1\%), detector efficiency (5\%), $\gamma$-intensities (3\%) and the individual uncertainty of the photo peak area determination (counting statistics :1-20\%).  

\begin{figure}
  \includegraphics[width=0.5\textwidth]{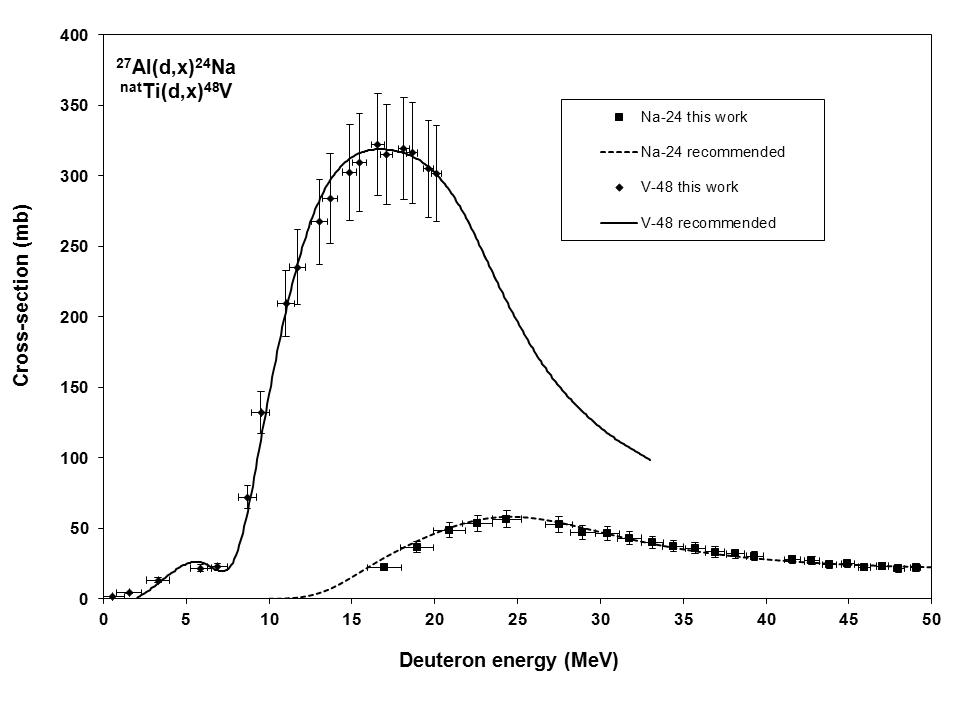}
\caption{The re-measured cross-sections of the used monitor reactions in comparison with the recommended data.}
\label{fig:1}       
\end{figure}

\begin{table}[t]
\tiny
\caption{\textbf{Decay characteristic of the investigated reaction products}}
\label{tab:1}       
\begin{tabular}{|p{0.45in}|p{0.3in}|p{0.3in}|p{0.25in}|p{0.55in}|p{0.4in}|} \hline 
Nuclide & Half-life & E$_{\gamma}$(keV) & I$_{\gamma}$(\%) & Contributing reaction & Q-value (keV) \\ \hline 
\textbf{$^{207}$Bi}\newline $\varepsilon $: 100~\% & 31.55 a & ~569.698\newline 1063.656 & ~97.75 \newline 74.5 & ${}^{206}$Pb(d,n)\newline ${}^{207}$Pb(d,2n)\newline ${}^{208}$Pb(d,3n)\newline  & 1333.45\newline -5404.32\newline -12772.19~ \\ \hline 
\textbf{${}^{206}$Bi\newline }$\varepsilon $: 100~\%\textbf{} & 6.243 d & ~183.977\newline ~343.51\newline 398.00\newline 497.06\newline 516.18\newline 537.45\newline 803.10\newline 881.01\newline 895.12\newline 1098.26 & ~15.8\newline ~23.5~\newline 10.75\newline 15.33\newline 40.8\newline 30.5\newline 99.0~\newline 66.2~\newline 15.67\newline ~13.51 & ${}^{206}$Pb(d,2n)\newline ${}^{207}$Pb(d,3n)\newline ${}^{208}$Pb(d,4n)\newline  & -6764.22\newline ~-13502.0\newline -20869.86 \\ \hline 
\textbf{${}^{205}$Bi\newline }~$\varepsilon $: 100~\%\textbf{\newline } & 15.31 d & 703.45\newline 987.66\newline ~1764.30\newline \newline  & 31.1\newline 16.1\newline ~32.5\newline  & $^{204}$Pb(d,n)\newline ${}^{206}$Pb(d,3n)\newline ${}^{207}$Pb(d,4n)\newline ${}^{208}$Pb(d,5n)\newline  & 1019.19\newline -13799.13\newline -20536.91\newline ~-27904.78\newline  \\ \hline 
\textbf{$^{204}$Bi\newline }$\varepsilon $: 99.75\%\newline $\beta^{+}$:0.25 \% & 11.22 h & 374.76\newline ~670.72\newline 899.15\newline 911.74\newline ~911.96\newline ~918.26\newline 983.98 & ~82\newline 11.4\newline 99\newline 13.6~\newline 11.2\newline ~ 10.9\newline 59~ & ${}^{204}$Pb(d,2n)\newline ${}^{206}$Pb(d,4n)\newline $^{207}$Pb(d,5n)\newline ${}^{208}$Pb(d,6n)\newline  & ~-7470.71\newline ~-22289.04\newline -29026.81\newline ~-36394.68 \\ \hline 
\textbf{${}^{203}$Bi\newline }$\varepsilon $: 100~\%\textbf{\newline } & 11.76 h & 816.3\newline 820.2\newline 825.2\newline 847.2\newline 896.9\newline 1033.7\newline 1679.6 & 4.1\newline 30.0\newline ~14.8\newline 8.6\newline 13.2\newline 8.9 \newline 8.9 & ${}^{204}$Pb(d,3n)\newline ${}^{206}$Pb(d,5n)\newline ${}^{207}$Pb(d,6n)\newline ${}^{208}$Pb(d,7n)\newline  & -14663.4\newline -29481.7\newline ~-36219.5\newline ~-43587.3 \\ \hline 
\textbf{$^{202}$Bi\newline }$\varepsilon $: 100~\%~\textbf{\newline } & 1.71 h & 422.13\newline 657.49\newline 960.67 & 83.7\newline 60.6\newline ~99.283 & ${}^{204}$Pb(d,4n)\newline ${}^{206}$Pb(d,6n)\newline ${}^{207}$Pb(d,7n)\newline ${}^{208}$Pb(d,8n)\newline  & ~-23518.1\newline -38336.5\newline -45074.2\newline -52442.1~~ \\ \hline 
\textbf{${}^{203}$Pb\newline }$\varepsilon $: 100~\%\textbf{} & 51.92 h & 279.1952 & 80.9 & ${}^{204}$Pb(d,p2n)\newline ${}^{206}$Pb(d,p4n)\newline ${}^{207}$Pb(d,p5n)\newline ${}^{208}$Pb(d,p6n)\newline ${}^{203}$Bi decay & -10619.24\newline -25437.57\newline ~-32175.35\newline ~-39543.22 \\ \hline 
\textbf{${}^{202m}$Pb\newline }$\varepsilon $: 9.5~\%\textbf{\newline }IT: 90.5~\%\newline 2169.83 keV\textbf{} & 3.54 h & 422.12\newline 657.49\newline 786.99\newline 960.70 & 84\newline 31.7\newline 49\newline 89.9~ & ${}^{204}$Pb(d,p3n)\newline ${}^{206}$Pb(d,p5n)\newline ${}^{207}$Pb(d,p6n)\newline ${}^{208}$Pb(d,p7n)\newline  & ~-17536.35\newline -32354.67\newline ~-39092.45\newline -46460.32 \\ \hline 
\textbf{$^{201}$Pb\newline }$\varepsilon \leq$100 \%\textbf{ ~\newline }$\beta^{+}$:0.064 \%\textbf{} & ~9.33 h & 331.15~\newline 907.67\newline 945.96 & 77~\newline 6.~7\newline 7.2~ & ${}^{204}$Pb(d,p4n)\newline ${}^{206}$Pb(d,p6n)\newline ${}^{207}$Pb(d,p7n)\newline ${}^{208}$Pb(d,p8n)\newline ${}^{20}$${}^{1}$Bi decay\newline  & ~-26288.5\newline ~-41106.8\newline -47844.6\newline -55212.5 \\ \hline 
\textbf{$^{202}$Tl\newline }$\varepsilon $: 100~\%~\textbf{} & ~12.31 d & 39.510~ & 91.5 & ${}^{204}$Pb(d,2p2n)\newline ${}^{206}$Pb(d,2p4n)\newline ${}^{207}$Pb(d,2p5n)\newline ${}^{208}$Pb(d,2p6n)\newline $^{202}$Pb decay & ~-16707.8\newline -31526.1\newline ~~-38263.9\newline -45631.8 \\ \hline 
\textbf{${}^{201}$Tl\newline }$\varepsilon $: 100~\%~\textbf{} & 3.0421 d & 135.34~\newline ~167.43 & 2.565\newline ~10.0 & ${}^{204}$Pb(d,2p3n)\newline ${}^{206}$Pb(d,2p5n)\newline ${}^{207}$Pb(d,2p6n)\newline ${}^{208}$Pb(d,2p7n)\newline ${}^{201}$Pb decay & ~-23586.6\newline ~~-38404.9\newline -45142.7\newline -52510.6~ \\ \hline 
\end{tabular}

\begin{flushleft}
\tiny{\noindent When complex particles are emitted instead of individual protons and neutrons the Q-values have to be modified by the respective binding energies of the compound particles: np-d, +2.2 MeV; 2np-t, +8.48 MeV; 2p2n-a, 28.30 MeV. 

\noindent *Decrease Q-values for isomeric states with level energy of the isomer }

\end{flushleft}
\end{table}

\section{Results and discussion}
\label{sec:3}

\subsection{Theoretical calculations}
\label{3.1}

For theoretical estimation the updated ALICE-IPPE-D \cite{13} and EMPIRE-D \cite{14} codes were used. In the modified  versions of the codes \cite{15,16} a simulation of direct (d,p) and (d,t) transitions by the general relations for a nucleon transfer probability in the continuum is included through an energy dependent enhancement factor for the corresponding transitions based on systematic of experimental data \cite{17}. Results of ALICE-IPPE-D for excited states were obtained by applying the isomeric ratios derived from the EMPIRE-D code to the total cross-sections calculated by ALICE-IPPE-D.
For a comparison with the experiments the theoretical data from the successive TENDL-2009-2010-2011-2012 and TENDL-2013 libraries \cite{18} (based on the modified TALYS code \cite{19} were used too to see the evolution of prediction capability and performance of the new versions.

\subsection{Excitation functions}
\label{3.2}

The cross-sections for radionuclides produced in the bombardment of $^{nat}$Pb with deuterons are tabulated in Tables 2 and 3 respectively, and are shown graphically in Figs 2-13 for comparison with the theory and with the earlier experimental results. By irradiating lead with natural isotopic abundance ($^{204}$Pb-1.4 \%, $^{206}$Pb-24.1 \%, $^{207}$Pb-22.1 \% and $^{208}$Pb-52.4 \%) with 50 MeV deuterons, radioisotopes of Bi, Pb and Tl are produced in significant amounts. We did not obtained reliable data for production of a few Bi, Pb and Tl radionuclides. Among the radio-products formed $^{208}$Bi ($T_{1/2}$ = 3.68*10$^5$5 a) and $^{205}$Pb  ($T_{1/2}$ = 1.73*10$^7$ a), $^{202g}$Pb ($T_{1/2}$ = 5.25*10$^3$ a)  have too long half-lives. The nuclides $^{209}$Pb and $^{204}$Tl have no $\gamma$-emission. In the case of the cumulative production of $^{201g}$Bi and $^{200}$Tl the parent nucleus or the internally decaying isomeric state has a half-life similar to that of the daughter state and separate assessment is not possible. Some of the produced radio-nuclei have half-lives that are too short compared to the experimental circumstances (long waiting time in high energy irradiation) resulting in activities at the moment of first measurement below the detection limits of our spectrometric set-up. 

\subsubsection{Bismuth isotopes}
\label{3.2.1}

The investigated radioisotopes of bismuth are produced only directly via $^{nat}$Pb(d,xn) reactions.

\vspace{0.3 cm}
\textbf{Production of  $^{207}$Bi}\\

The reaction cross-sections of $^{207}$Bi ($T_{1/2}$ = 31.55 a)  are shown in Fig 2, together with the earlier experimental results and theoretical estimations. The new experimental data are higher compared to Ditroi et al. \cite{3} and are in good agreement with the results of Wasilyevski et al. \cite{4}.  By comparing with theoretical results disagreements in the maximum and in the high energy pre-compound tail can be seen. The earlier versions of the TENDL libraries (From 2010 back) differ significantly from the experimental data. TENDL-2012 and 2013 seems to be identical. EMPIRE-D strongly overestimates the experimental data, while ALICE-IPPE seems to be better than the modified ALICE-IPPE-D.

\begin{figure}
  \includegraphics[width=0.5\textwidth]{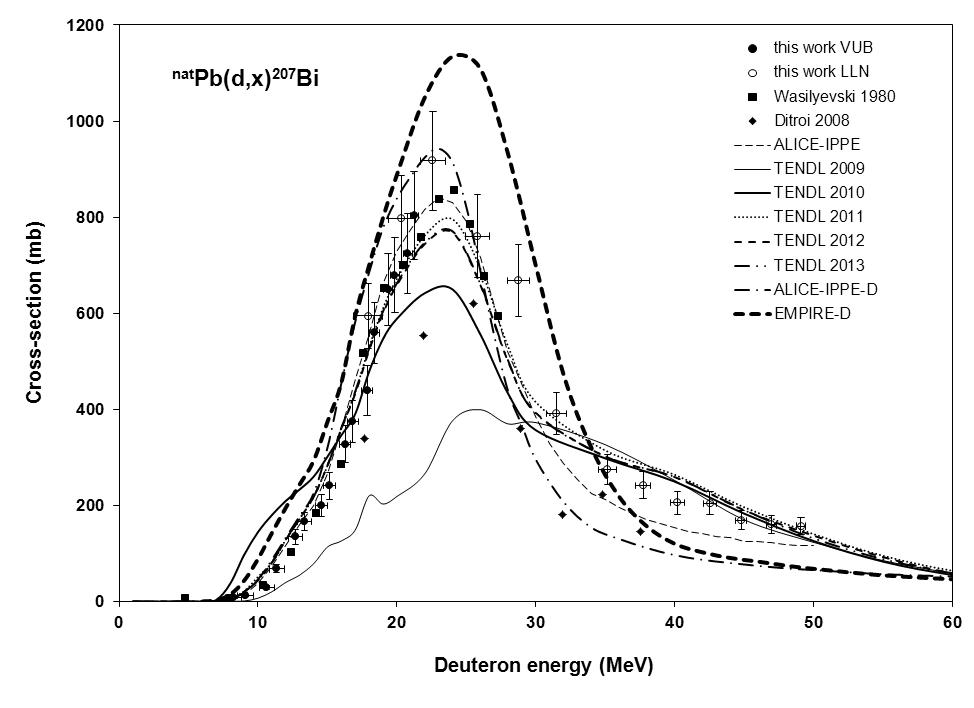}
\caption{Excitation function of the $^{nat}$Pb(d,x)$^{207}$Bi reaction compared with the theory and literature}
\label{fig:2}       
\end{figure}

\vspace{0.3 cm}
\textbf{Production of  $^{206}$Bi}\\

The agreement between the different sets of experimental production cross-section data for $^{206}$Pb ($T_{1/2}$ = 6.243 d) is acceptable, except the energy region around 20 MeV (see Fig. 3). The theoretical results show large disagreement both in shape and in the magnitude. Especially TENDL-2012 shows large discrepancies compared to both TENDL-2011 and TENDL-2013. The ALICE results also overestimate the experimental values, the worst behavior is observed by EMPIRE.

\begin{figure}
  \includegraphics[width=0.5\textwidth]{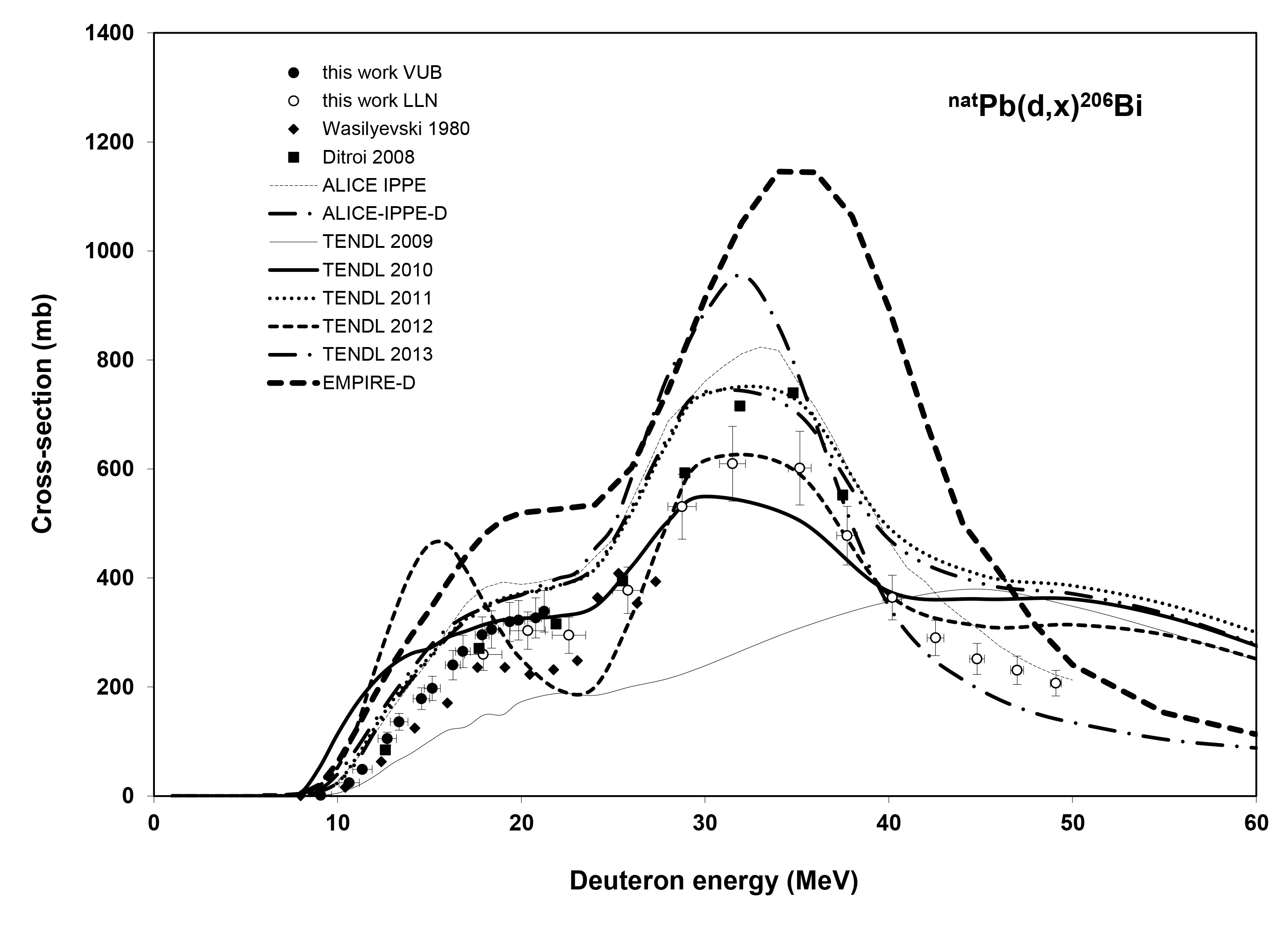}
\caption{Excitation function of the $^{nat}$Pb(d,x)$^{206}$Bi reaction compared with the theory and literature}
\label{fig:3}       
\end{figure}

\vspace{0.3 cm}
\textbf{Production of  $^{205}$Bi}\\

For $^{205}$Bi ($T_{1/2}$ = 15.31 d) there are large differences between the present experimental data  and the results  of Wasiljevski et al. \cite{4}, while a reasonable agreement is seen with Ditroi et al. \cite{3}. The ALICE-IPPE-D and the EMPIRE-D code results overestimate the presently measured data (Fig. 4). The improvement of the different versions of the TENDL libraries is well demonstrated in this figure.

\begin{figure}
  \includegraphics[width=0.5\textwidth]{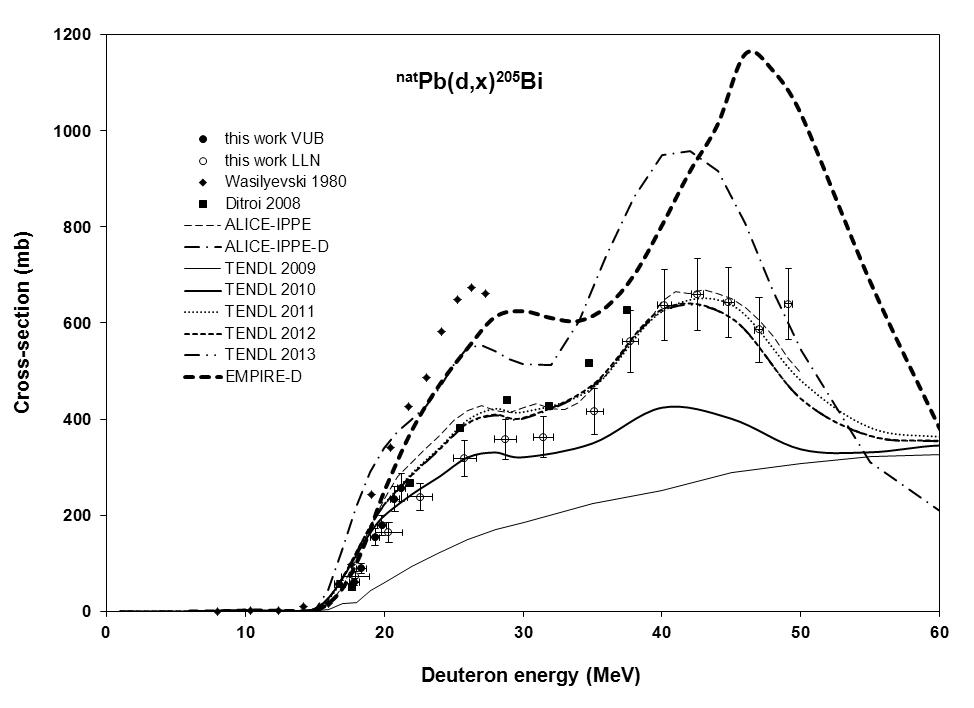}
\caption{Excitation function of the $^{nat}$Pb(d,x)$^{205}$Bi reaction compared with the theory and literature}
\label{fig:4}       
\end{figure}

\vspace{0.3 cm}
\textbf{Production of  $^{204}$Bi}\\

The experimental data for production of $^{204}$Bi ($T_{1/2}$ = 11.22 h) according to Fig. 5 show good agreement with the theoretical data in the latest TENDL libraries and the older ALICE-IPPE calculations. In this case the newest TENDL-2013 seems to be worse than the previous two above 40 MeV. The predictions of the ALICE-D and the EMPIRE-D are systematically high, especially in the energy region above 25 MeV. 

\begin{figure}
  \includegraphics[width=0.5\textwidth]{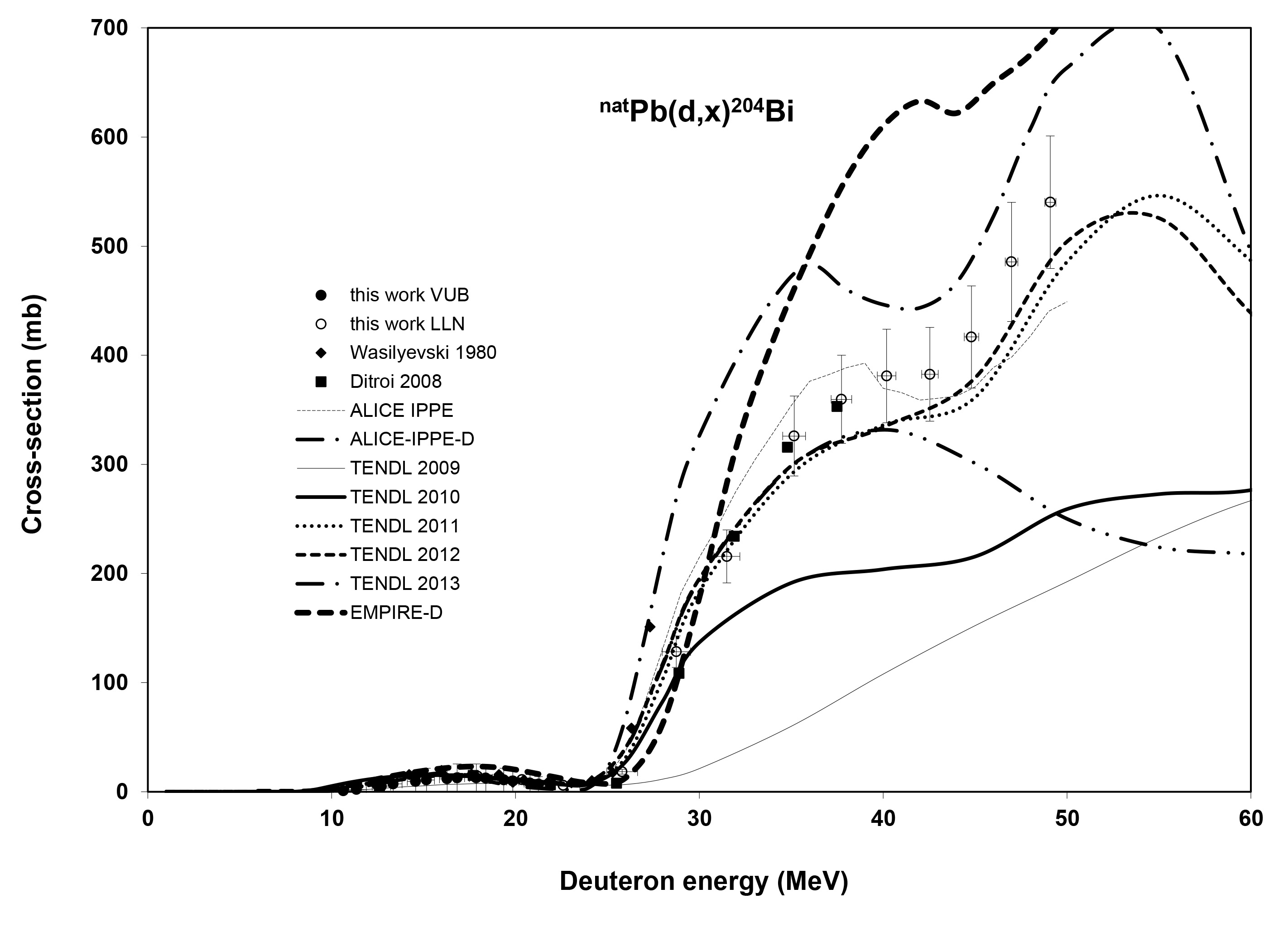}
\caption{Excitation function of the $^{nat}$Pb(d,x)$^{204}$Bi reaction compared with the theory and literature}
\label{fig:5}       
\end{figure}

\vspace{0.3 cm}
\textbf{Production of  $^{203}$Bi}\\

The different sets of experimental activation cross-sections of the $^{203}$Bi ($T_{1/2}$ = 11.76 h) show good agreement with the latest three TENDL data (Fig. 6). The predictions of the ALICE-D and the EMPIRE-D are systematically high. 

\begin{figure}
  \includegraphics[width=0.5\textwidth]{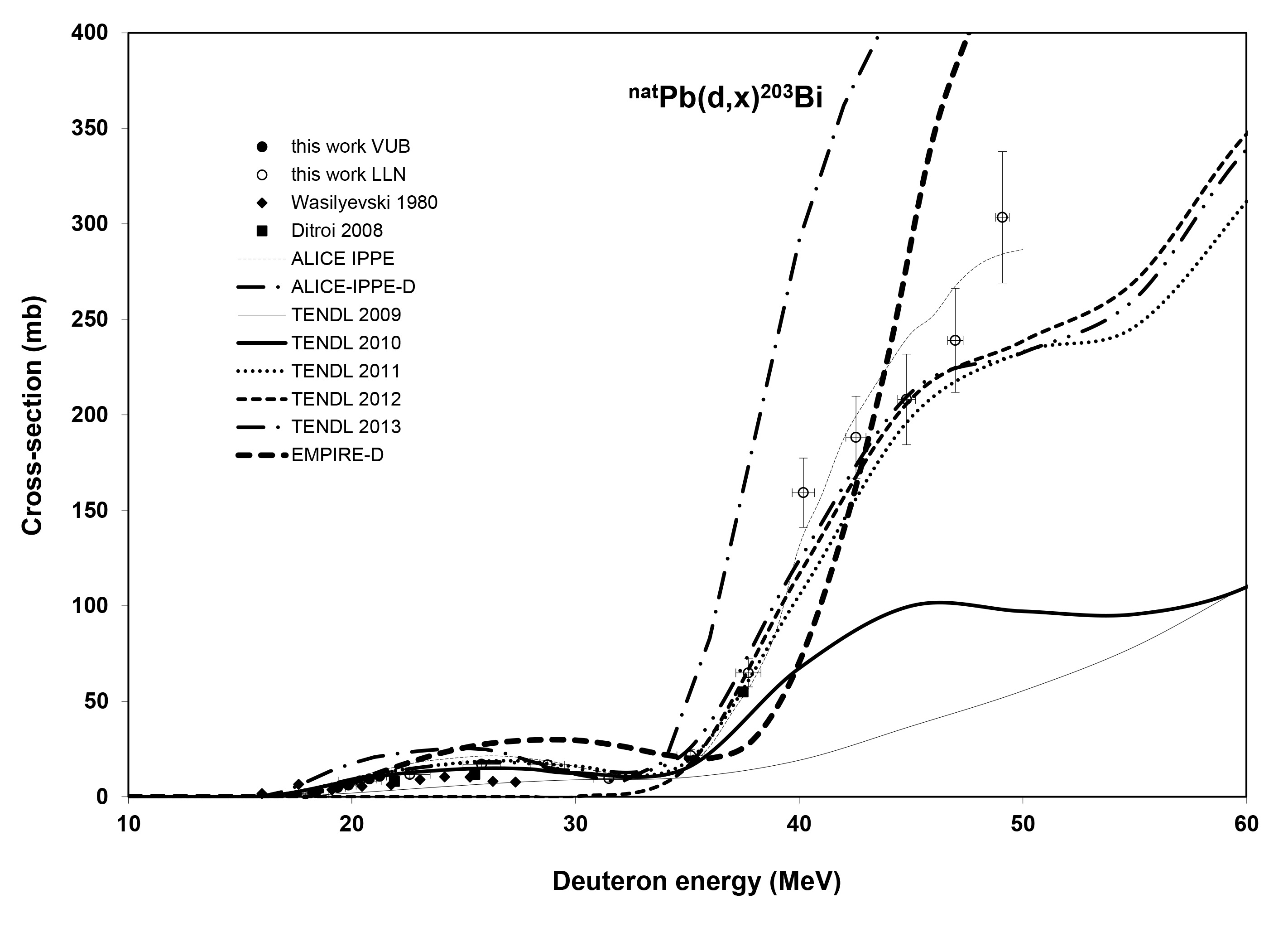}
\caption{Excitation function of the $^{nat}$Pb(d,x)$^{203}$Bi reaction compared with the theory and literature}
\label{fig:6}       
\end{figure}

\vspace{0.3 cm}
\textbf{Production of  $^{202}$Bi}\\

No earlier experimental data were found. The agreement of our experimental data with the theoretical predictions for production of $^{202}$Bi  ($T_{1/2}$ = 1.71 h) is the best in case of TENDL-2012 (see Fig. 7).

\begin{figure}
  \includegraphics[width=0.5\textwidth]{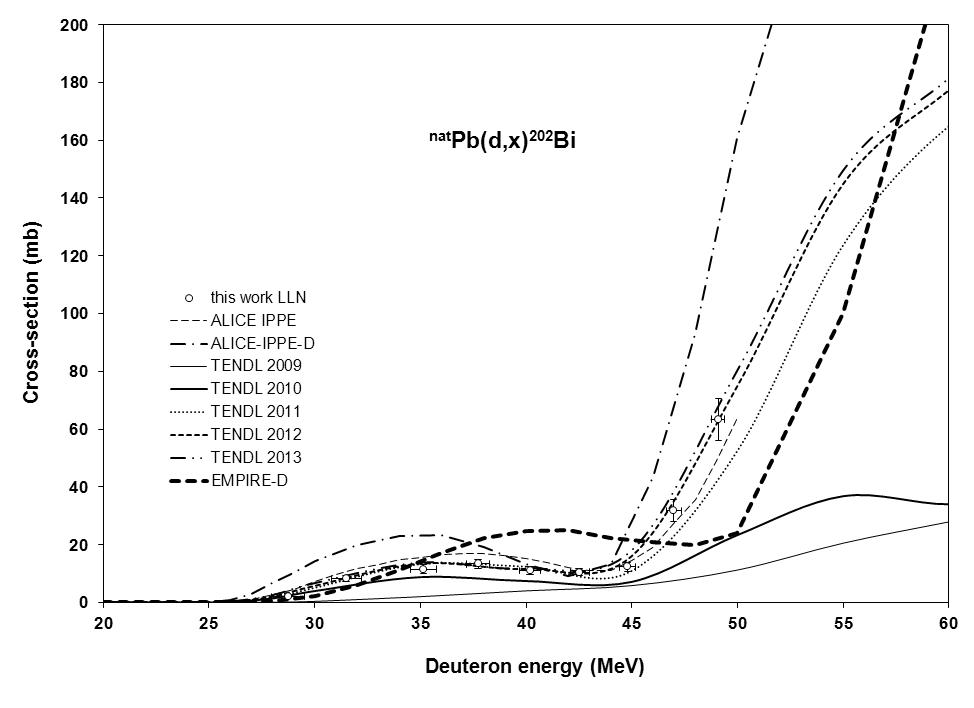}
\caption{Excitation function of the $^{nat}$Pb(d,x)$^{202}$Bi reaction compared with the theory}
\label{fig:7}       
\end{figure}

\subsubsection{Lead isotopes}
\label{3.2.2}

The investigated radioisotopes of lead are produced directly via (d,pxn) reactions and through EC, ?+ decay of the parent bismuth radioisotopes.

\vspace{0.3 cm}
\textbf{Production of  $^{209}$Pb}\\

As it was mentioned earlier $^{209}$Pb produced via the $^{208}$Pb(d,p) reaction ($T_{1/2}$ = 3.253 h, $\beta^-$: 100 \%) was not detected in this experiment due to lack of ?-lines. In several of our earlier investigations we obtained experimental results for (d,p) reactions that were used to develop and implement improvements for the description of the (d,p) process in the model codes. Therefore in this work we made new theoretical calculation for this reaction using the upgraded ALICE-IPPE-D and EMPIRE-D codes and compared these results with the experimental data available in the literature ($\gamma$- measurements), with the excitation function based on experimental data of the (d,p) reaction in this mass region ("systematics") and with results of the TALYS code in the TENDL libraries. As it is shown in Fig. 8, there is acceptable good agreement of the literature experimental data for $^{209}$Pb production with the systematics and with ALICE-IPPE-D results. The EMPIRE-D and TENDL data significantly underestimate the maximum.

\begin{figure}
  \includegraphics[width=0.5\textwidth]{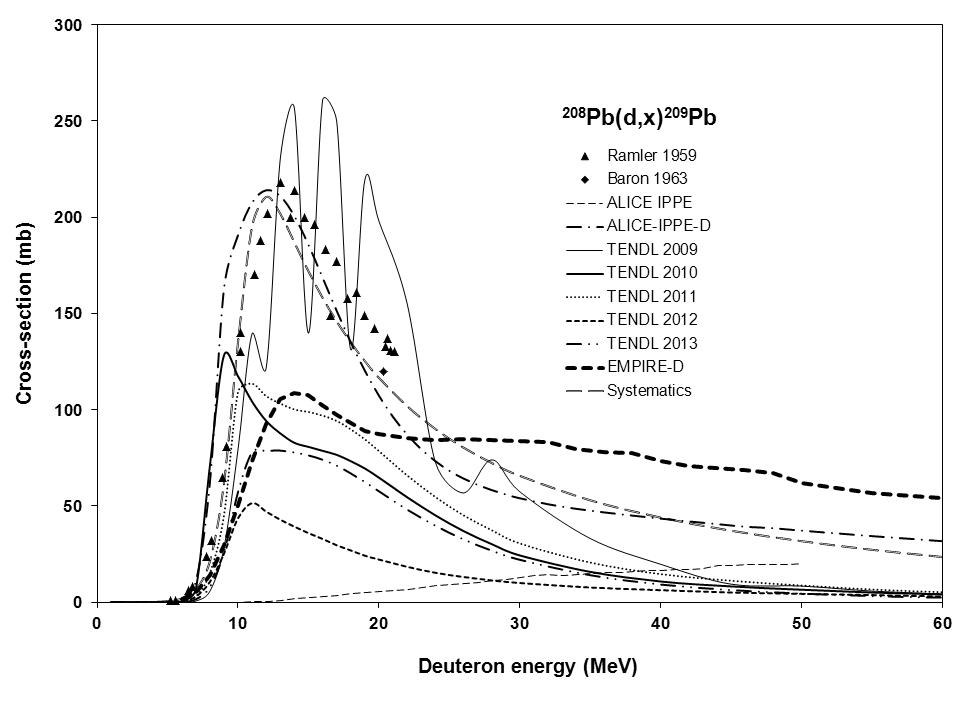}
\caption{Excitation function of the $^{nat}$Pb(d,x)$^{209}$Pb reaction calculated by the theory and literature data}
\label{fig:8}       
\end{figure}

\vspace{0.3 cm}
\textbf{Production of  $^{203g}$Pb}\\

The radioisotope $^{203g}$Pb ($T_{1/2}$ = 51.92 h) is produced directly, through the internal decay of the short half-life isomeric state (6.21 s) and from the decay of the $^{203}$Bi ($T_{1/2}$ = 11.76 h) parent. In Fig. 9 we present experimental data for the cumulative production, measured after complete decay of all states contributing indirectly, in comparison with the earlier result of Ditroi et al. \cite{3} and with the results of the theoretical calculations. Best agreement was found in this case with the EMPIRE-D results.  The last three versions of TENDL run almost together.

\begin{figure}
  \includegraphics[width=0.5\textwidth]{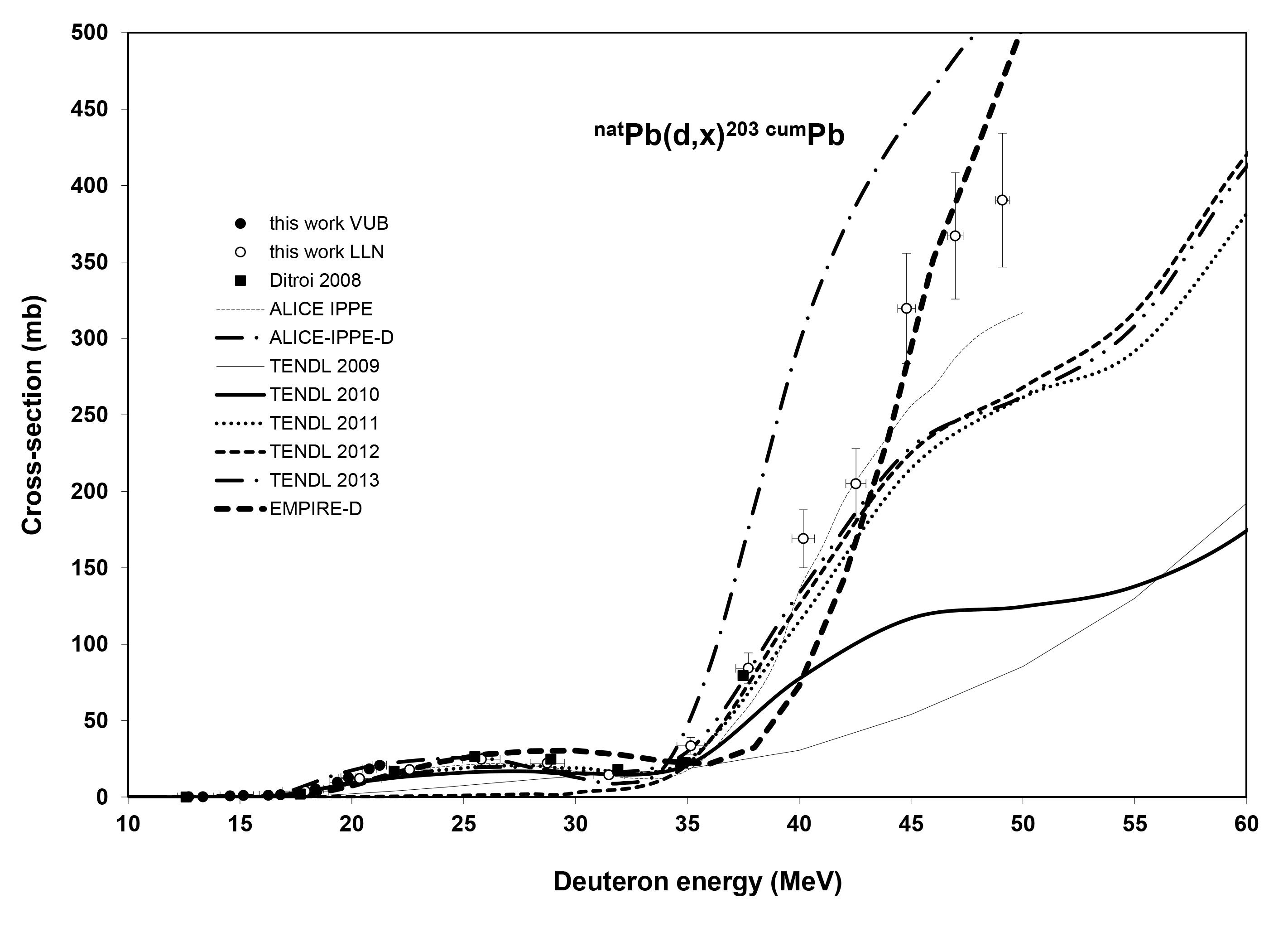}
\caption{Excitation function of the $^{nat}$Pb(d,x)$^{203g}$Pb (cum) reaction compared with the theory and literature}
\label{fig:9}       
\end{figure}

\vspace{0.3 cm}
\textbf{Production of  $^{202m}$Pb}\\

Out of the two isomeric states of $^{202}$Pb the ground state has a very long half-life (5.25*10$^4$ a) and the decay is not followed by $\gamma$-emission. We could measure the activation cross-sections only for production of the shorter-lived (3.54 h) isomeric state. This state is produced only directly, as $^{202}$Bi ($T_{1/2}$ = 1.71 h) decays only to the ground state of $^{202}$Pb. No earlier experimental data were found in the literature. The experimental and the theoretical excitation functions are shown in Fig. 10.  The theoretical results show large disagreement in the investigated energy range.

\begin{figure}
  \includegraphics[width=0.5\textwidth]{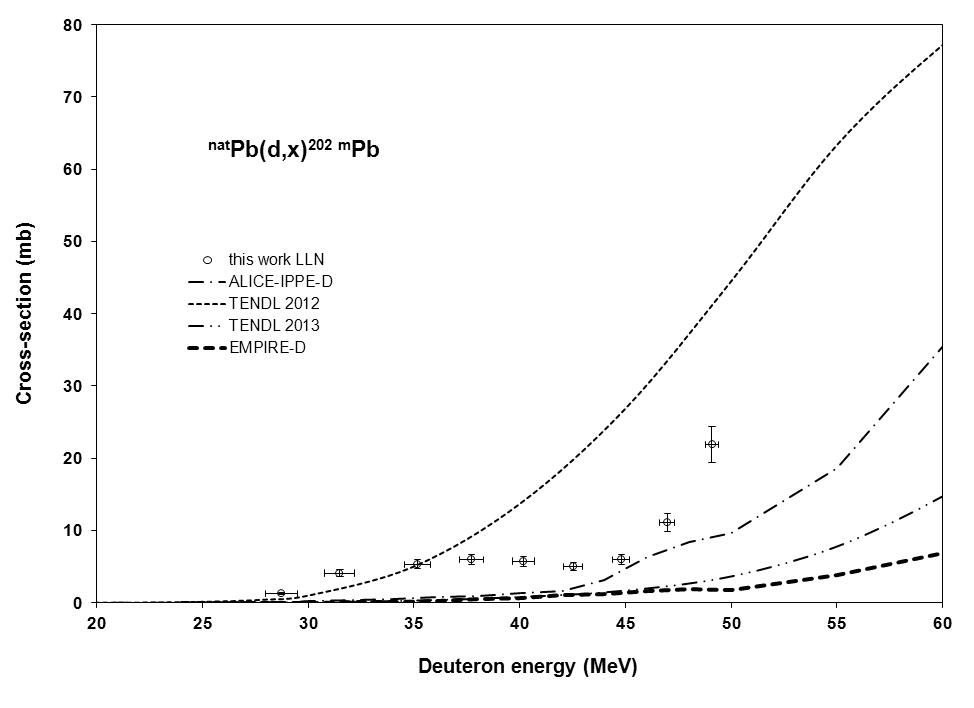}
\caption{Excitation function of the $^{nat}$Pb(d,x)$^{202m}$Pb reaction compared with the theory and literature}
\label{fig:10}       
\end{figure}

\vspace{0.3 cm}
\textbf{Production of  $^{201}$Pb}\\

The activation cross-sections for cumulative production of $^{201}$Pb  ground state ($T_{1/2}$ = 9.33 h) were measured after the decay of metastable  $^{201m}$Pb ($T_{1/2}$ = 61 s) and of both isomeric states of the parent $^{201}$Bi ($T_{1/2}$ = 59.1 min and 103 min respectively). The experimental and theoretical data are shown in Fig. 11. There is no significant difference between the last 3 TENDL versions up to 50 MeV and those give the best approximation with a little overestimation above 45 MeV.

\begin{figure}
  \includegraphics[width=0.5\textwidth]{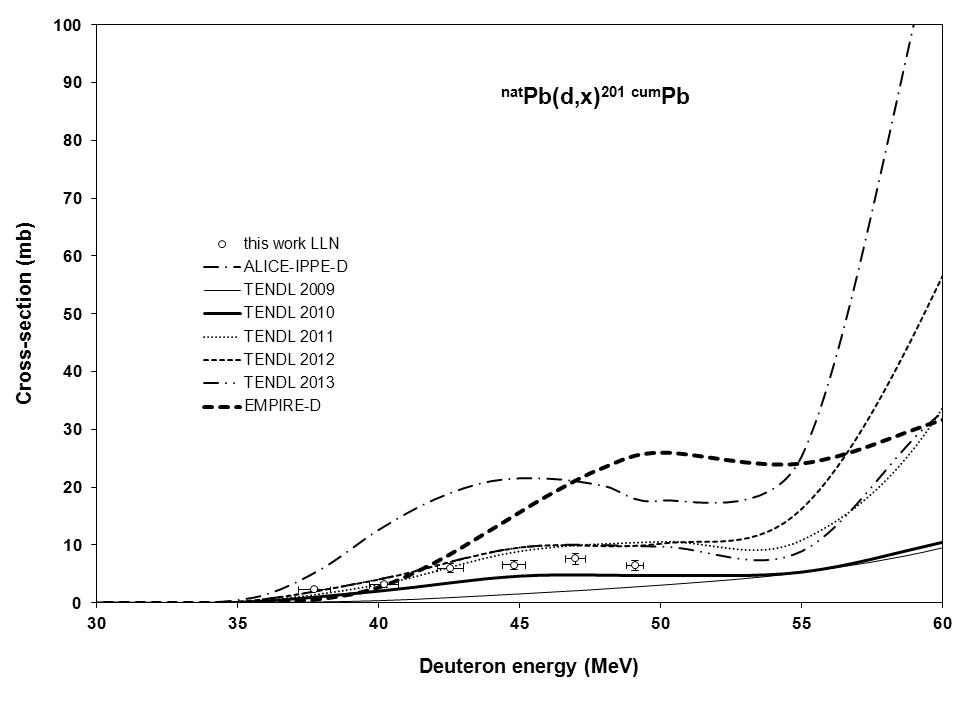}
\caption{Excitation function of the $^{nat}$Pb(d,x)$^{201}$Pb(cum) reaction compared with the theory and literature}
\label{fig:11}       
\end{figure}

\subsubsection{Thallium isotopes}
\label{3.2.3}

The measured radioisotopes of thallium are produced directly via (d,2pxn) reactions and through EC decay of parent Pb radioisotopes. 

\vspace{0.3 cm}
\textbf{Production of  $^{202}$Tl}\\

The cumulative activation cross-sections of $^{202}$Tl ($T_{1/2}$ = 12.31 d) contain the direct production and the decay of $^{202m}$Pb ($T_{1/2}$ = 3.54 h). The contribution from the decay of $^{202g}$Pb is negligible due to its very long half-life ($T_{1/2}$ = 5.25*10$^4$ a). The new data are lower compared to our earlier results (Ditroi et al. \cite{3}) (Fig. 12).  No theoretical results for the $^{202m}$Pb in the earlier TENDL libraries, therefore only the TENDL-2012 and TENDL-2013 data are shown. None of the codes describes reasonably the cumulative production.

\begin{figure}
  \includegraphics[width=0.5\textwidth]{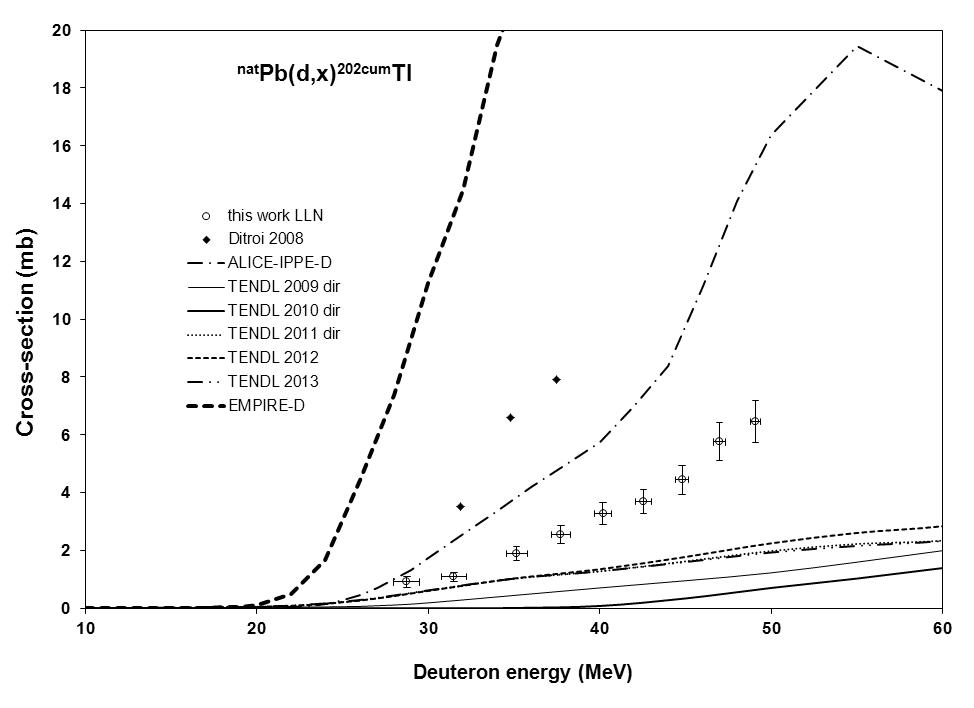}
\caption{Excitation function of the $^{nat}$Pb(d,x)$^{202}$Tl(cum) reaction compared with the theory and literature}
\label{fig:12}       
\end{figure}

\vspace{0.3 cm}
\textbf{Production of  $^{201}$Tl}\\

The measured production cross-sections of $^{201}$Tl ($T_{1/2}$ = 3.0421 d) are cumulative and include the decay chain contributions of the shorter-lived parent $^{201}$Bi ($T_{1/2}$ =5 9.1 min and 1.8 h), and $^{201}$Pb ($T_{1/2}$ = 9.33 h and 61 s) radioisotopes. The measured cross-sections are shown in Fig. 13. No earlier experimental data were found. Because of the only few measured points it is difficult to judge between the theoretical model calculation curves.

\begin{figure}
  \includegraphics[width=0.5\textwidth]{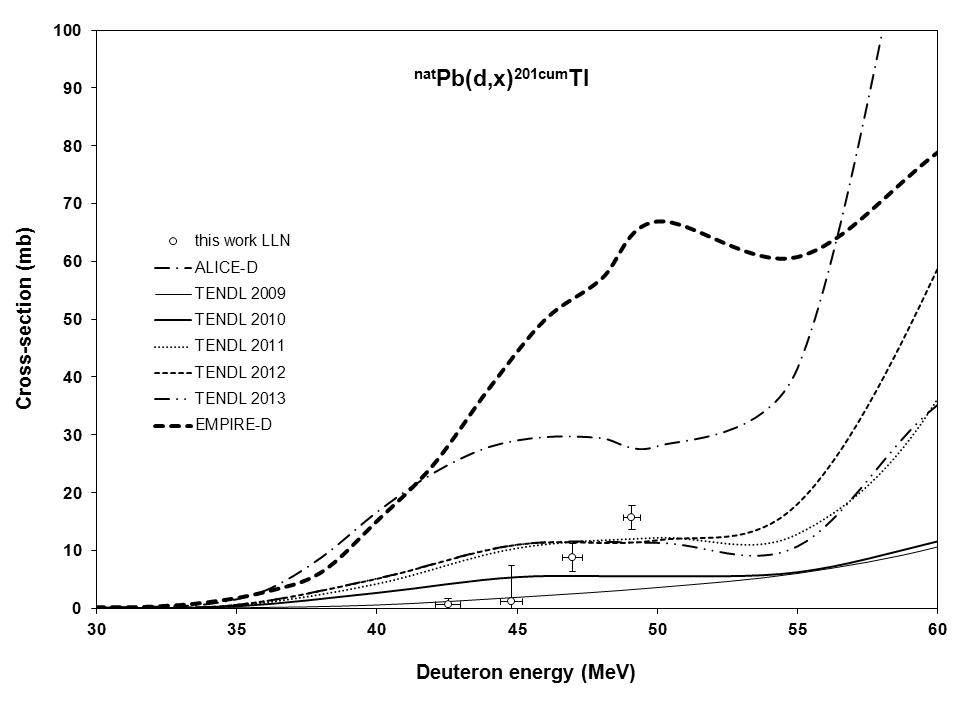}
\caption{Excitation function of the $^{nat}$Pb(d,x)$^{201}$Tl(cum) reaction compared with the theory and literature}
\label{fig:13}       
\end{figure}

\begin{table*}[t]
\tiny
\caption{\textbf{Measured cross-sections of the bismuth isotopes in LNL and VUB laboratories}}
\label{tab:2}       
\begin{tabular}{|c|c|c|c|c|c|c|c|c|c|c|c|c|c|c|} \hline 
\textbf{Lab.} & \multicolumn{2}{|p{0.8in}|}{\textbf{Energy\newline }E+$\Delta$DE(MeV) } & \multicolumn{2}{|p{0.8in}|}{\textbf{${}^{207}$Bi \newline }$\sigma\pm\Delta\sigma$(mb)} & \multicolumn{2}{|p{0.8in}|}{\textbf{${}^{206}$Bi \newline }$\sigma\pm\Delta\sigma$(mb)} & \multicolumn{2}{|p{0.8in}|}{\textbf{${}^{205}$Bi \newline }$\sigma\pm\Delta\sigma$(mb)} & \multicolumn{2}{|p{0.8in}|}{\textbf{${}^{204}$Bi \newline }$\sigma\pm\Delta\sigma$(mb)} & \multicolumn{2}{|p{0.8in}|}{\textbf{${}^{203}$Bi \newline }$\sigma\pm\Delta\sigma$(mb)} & \multicolumn{2}{|p{0.8in}|}{\textbf{${}^{202}$Bi \newline }$\sigma\pm\Delta\sigma$(mb)} \\ \hline 
LNL & 17.9 & 1.0 & 594.1 & 67.0 & 259.7 & 29.2 & 72.5 & 9.5 & 14.2 & 1.7 & 1.5 & 0.4 & ~ &  \\ \hline 
 & 20.4 & 1.0 & 796.1 & 89.6 & 303.4 & 34.2 & 164.5 & 20.6 & 11.1 & 1.3 & 8.2 & 1.3 & ~ &  \\ \hline 
 & 22.6 & 0.9 & 916.8 & 103.1 & 295.0 & 33.2 & 238.0 & 28.9 & 5.9 & 0.8 & 11.7 & 1.8 & ~ &  \\ \hline 
 & 25.8 & 0.8 & 760.5 & 86.3 & 377.4 & 42.4 & 318.9 & 37.4 & 18.3 & 2.1 & 17.0 & 2.2 & ~ &  \\ \hline 
 & 28.7 & 0.8 & 668.3 & 75.3 & 531.0 & 59.7 & 358.4 & 41.6 & 128.3 & 14.5 & 16.7 & 2.3 & 2.0 & 0.4 \\ \hline 
 & 31.5 & 0.7 & 390.6 & 44.0 & 609.6 & 68.5 & 362.5 & 42.0 & 215.6 & 24.3 & 9.6 & 1.8 & 8.3 & 1.0 \\ \hline 
 & 35.2 & 0.6 & 274.0 & 31.0 & 601.6 & 67.6 & 415.8 & 48.1 & 326.0 & 36.7 & 21.5 & 3.2 & 11.5 & 1.5 \\ \hline 
 & 37.7 & 0.6 & 241.5 & 28.2 & 477.6 & 53.7 & 561.3 & 64.3 & 359.6 & 40.4 & 64.9 & 7.4 & 13.3 & 1.6 \\ \hline 
 & 40.2 & 0.5 & 205.4 & 23.7 & 364.0 & 41.0 & 637.2 & 73.2 & 381.1 & 42.9 & 159.2 & 18.2 & 11.1 & 1.5 \\ \hline 
 & 42.5 & 0.5 & 204.1 & 23.9 & 290.0 & 32.6 & 659.8 & 75.3 & 382.6 & 43.0 & 188.2 & 21.5 & 10.5 & 1.3 \\ \hline 
 & 44.8 & 0.4 & 168.7 & 19.9 & 251.5 & 28.3 & 642.3 & 73.5 & 416.8 & 46.9 & 208.1 & 23.7 & 12.3 & 1.5 \\ \hline 
 & 47.0 & 0.3 & 159.2 & 18.5 & 230.9 & 26.0 & 585.6 & 67.2 & 485.6 & 54.6 & 238.9 & 27.2 & 31.8 & 3.7 \\ \hline 
 & 49.1 & 0.3 & 156.8 & 18.4 & 207.0 & 23.3 & 639.4 & 73.1 & 540.2 & 60.7 & 303.4 & 34.4 & 63.4 & 7.2 \\ \hline 
 &  &  &  &  &  &  &  &  &  &  &  &  &  &  \\ \hline 
VUB & 8.2 & 0.6 & 7.5 & 1.6 & ~ &  & ~ &  & ~ &  & ~ &  &  &  \\ \hline 
 & 9.1 & 0.6 & 12.7 & 2.1 & 1.4 & 0.2 & ~ &  & ~ &  & ~ &  &  &  \\ \hline 
 & 10.6 & 0.5 & 29.0 & 4.5 & 24.2 & 2.7 & ~ &  & 0.9 & 0.1 & ~ &  &  &  \\ \hline 
 & 11.4 & 0.5 & 67.7 & 8.4 & 48.9 & 5.5 & ~ &  & 2.1 & 0.2 & ~ &  &  &  \\ \hline 
 & 12.7 & 0.5 & 134.2 & 15.7 & 105.4 & 11.8 & ~ &  & 5.0 & 0.6 & ~ &  &  &  \\ \hline 
 & 13.3 & 0.5 & 166.3 & 19.1 & 136.1 & 15.3 & ~ &  & 7.1 & 0.8 & ~ &  &  &  \\ \hline 
 & 14.6 & 0.5 & 200.0 & 23.1 & 178.5 & 20.1 & ~ &  & 9.4 & 1.1 & ~ &  &  &  \\ \hline 
 & 15.2 & 0.4 & 240.8 & 28.3 & 197.4 & 22.2 & ~ &  & 10.7 & 1.2 & ~ &  &  &  \\ \hline 
 & 16.3 & 0.4 & 327.3 & 38.3 & 240.2 & 27.0 & ~ &  & 11.7 & 1.3 & ~ &  &  &  \\ \hline 
 & 16.8 & 0.4 & 373.9 & 43.8 & 265.2 & 29.8 & 56.9 & 7.0 & 12.7 & 1.4 & ~ &  &  &  \\ \hline 
 & 17.9 & 0.4 & 438.7 & 51.7 & 295.4 & 33.2 & 60.8 & 7.4 & 12.7 & 1.4 & ~ &  &  &  \\ \hline 
 & 18.4 & 0.4 & 558.8 & 64.1 & 305.5 & 34.3 & 89.0 & 10.5 & 12.4 & 1.4 & 2.2 & 0.3 &  &  \\ \hline 
 & 19.4 & 0.3 & 648.9 & 75.0 & 319.5 & 35.9 & 153.8 & 17.7 & 10.7 & 1.2 & 4.9 & 0.6 &  &  \\ \hline 
 & 19.9 & 0.3 & 679.4 & 78.1 & 322.4 & 36.2 & 179.1 & 20.5 & 9.7 & 1.1 & 6.1 & 0.7 &  &  \\ \hline 
 & 20.8 & 0.3 & 724.6 & 83.2 & 326.8 & 36.7 & 233.4 & 26.5 & 7.8 & 0.9 & 9.3 & 1.1 &  &  \\ \hline 
 & 21.2 & 0.3 & 803.0 & 92.2 & 338.6 & 38.0 & 256.7 & 29.1 & 6.9 & 0.8 & 10.8 & 1.2 &  &  \\ \hline 
\end{tabular}

\end{table*}

\begin{table*}[t]
\tiny
\caption{\textbf{Measured cross-sections of the bismuth isotopes in LNL and VUB laboratories}}
\label{tab:3}       
\begin{tabular}{|c|c|c|c|c|c|c|c|c|c|c|c|c|} \hline 
\textbf{Lab.} & \multicolumn{2}{|p{0.8in}|}{\textbf{Energy\newline }E+$\Delta$E(MeV)} & \multicolumn{2}{|p{0.8in}|}{\textbf{${}^{203}$Pb \newline }$\sigma\pm\Delta\sigma$(mb)} & \multicolumn{2}{|p{0.8in}|}{\textbf{${}^{202m}$Pb \newline }$\sigma\pm\Delta\sigma$(mb)} & \multicolumn{2}{|p{0.8in}|}{\textbf{${}^{201}$Pb \newline }$\sigma\pm\Delta\sigma$(mb)} & \multicolumn{2}{|p{0.8in}|}{\textbf{${}^{202}$Tl \newline }$\sigma\pm\Delta\sigma$(mb)} & \multicolumn{2}{|c|}{\textbf{${}^{201}$Tl \newline }$\sigma\pm\Delta\sigma$(mb)} \\ \hline 
LNL & 17.9 & 1.0 & 3.9 & 0.6 & ~ &  & ~ &  & ~ &  & ~ &  \\ \hline 
 & 20.4 & 1.0 & 12.1 & 1.4 & ~ &  & ~ &  & ~ &  & ~ &  \\ \hline 
 & 22.6 & 0.9 & 17.9 & 2.0 & ~ &  & ~ &  & ~ &  & ~ &  \\ \hline 
 & 25.8 & 0.8 & 24.8 & 4.1 & ~ &  & ~ &  & ~ &  & ~ &  \\ \hline 
 & 28.7 & 0.8 & 22.2 & 2.5 & 1.3 & 0.2 & ~ &  & 0.9 & 0.2 & ~ &  \\ \hline 
 & 31.5 & 0.7 & 14.6 & 1.7 & 4.1 & 0.5 & ~ &  & 1.1 & 0.2 & ~ &  \\ \hline 
 & 35.2 & 0.6 & 33.5 & 5.5 & 5.3 & 0.6 & ~ &  & 1.9 & 0.2 & ~ &  \\ \hline 
 & 37.7 & 0.6 & 84.2 & 10.1 & 6.0 & 0.7 & 2.3 & 0.3 & 2.5 & 0.3 & ~ &  \\ \hline 
 & 40.2 & 0.5 & 169.0 & 19.0 & 5.7 & 0.7 & 3.1 & 0.4 & 3.3 & 0.4 & ~ &  \\ \hline 
 & 42.5 & 0.5 & 205.0 & 23.0 & 5.0 & 0.6 & 6.0 & 0.7 & 3.7 & 0.4 & 0.6 & 1.0 \\ \hline 
 & 44.8 & 0.4 & 319.6 & 36.1 & 5.9 & 0.7 & 6.5 & 0.8 & 4.4 & 0.5 & 1.1 & 6.3 \\ \hline 
 & 47.0 & 0.3 & 367.1 & 41.3 & 11.0 & 1.3 & 7.6 & 0.9 & 5.8 & 0.7 & 8.8 & 2.5 \\ \hline 
 & 49.1 & 0.3 & 390.4 & 43.8 & 21.8 & 2.5 & 6.4 & 0.8 & 6.5 & 0.7 & 15.7 & 2.1 \\ \hline 
 &  &  &  &  &  &  &  &  &  &  &  &  \\ \hline 
VUB & 12.7 & 0.5 & 0.09 & 0.21 &  &  &  &  &  &  &  &  \\ \hline 
 & 13.3 & 0.5 & 0.14 & 0.13 &  &  &  &  &  &  &  &  \\ \hline 
 & 14.6 & 0.5 & 0.79 & 0.15 &  &  &  &  &  &  &  &  \\ \hline 
 & 15.2 & 0.4 & 1.1 & 0.17 &  &  &  &  &  &  &  &  \\ \hline 
 & 16.3 & 0.4 & 1.1 & 0.17 &  &  &  &  &  &  &  &  \\ \hline 
 & 16.8 & 0.4 & 1.4 & 0.20 &  &  &  &  &  &  &  &  \\ \hline 
 & 17.9 & 0.4 & 3.4 & 0.42 &  &  &  &  &  &  &  &  \\ \hline 
 & 18.4 & 0.4 & 5.3 & 0.62 &  &  &  &  &  &  &  &  \\ \hline 
 & 19.4 & 0.3 & 9.6 & 1.1 &  &  &  &  &  &  &  &  \\ \hline 
 & 19.9 & 0.3 & 12.8 & 1.5 &  &  &  &  &  &  &  &  \\ \hline 
 & 20.8 & 0.3 & 18.4 & 2.1 &  &  &  &  &  &  &  &  \\ \hline 
 & 21.2 & 0.3 & 20.8 & 2.3 &  &  &  &  &  &  &  &  \\ \hline 

\end{tabular}
\end{table*}

\subsection{Production yields}
\label{3.3}

By fitting and integrating the measured excitation functions we have deduced integral yields for the investigated reactions (Fig. 14-15).  The calculated values of $^{207,206,205}$Bi were compared with the experimental thick target yields available in the literature (Dmitirev et al. \cite{5}) (Fig. 14). The yields of the rest of the measured isotopes are presented in Fig. 15. No literature data were found for these isotopes. In Fig. 14 t is seen that the larges yield can be achieved by $^{206}$Bi the previous results of Dmitriev \cite{5} is in good agreement with our results on $^{207}$Bi and somewhat higher in the case of $^{206}$Bi and $^{205}$Bi. In Fig. 15 much larger yield are also presented (e.g. $^{204}$Bi) and the range of yield covers even four orders of magnitude by 30 MeV. 

\begin{figure}
  \includegraphics[width=0.5\textwidth]{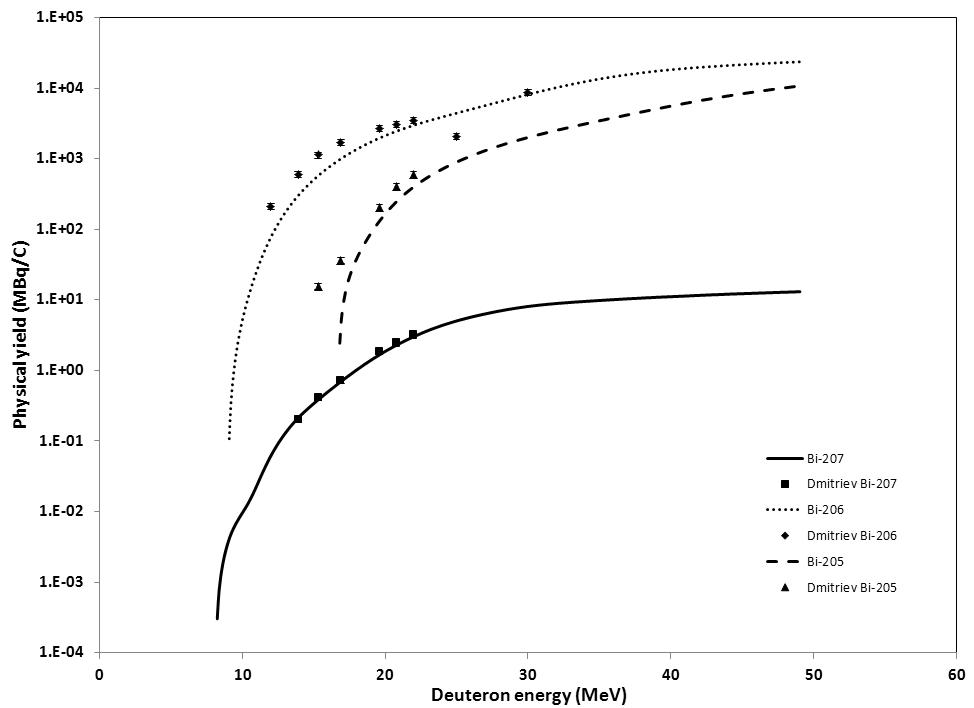}
\caption{Calculated integral yields of  $^{205,206,207}$Bi on natural lead target, compared with the literature}
\label{fig:14}       
\end{figure}

\begin{figure}
  \includegraphics[width=0.5\textwidth]{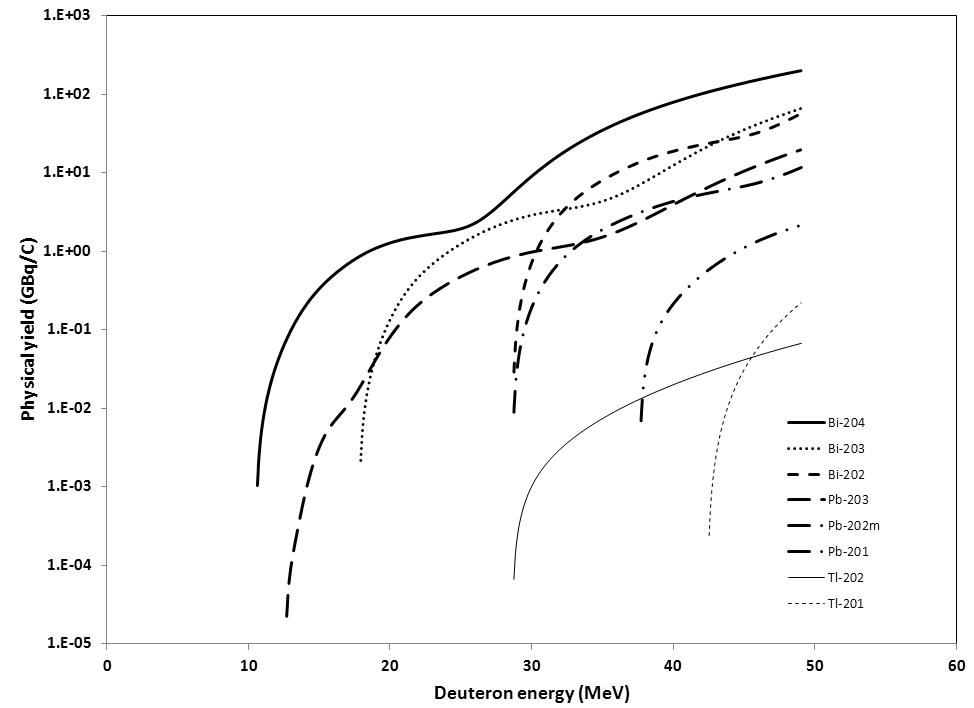}
\caption{15.	Calculated integral yields of  $^{204,203,202}$Bi, $^{203,202m,201}$Pb and $^{202,201}$Tl on natural lead target}
\label{fig:15}       
\end{figure}

\section{Summary and conclusions}
\label{sec:4}
Lead is a metallic material, which is frequently used as construction material and alloy component in industry and in nuclear technology, including accelerator technology (fusion, shielding and target technology, accelerator driven waste transmutation, etc.). Experimental cross-sections and derived integral yields are reported for the $^{nat}$Pb(d,x)$^{206,205,204,203,202}$Bi, $^{203cum,202m,201cum}$Pb and $^{202cum,201cum}$Tl reactions. The comparison with the earlier experimental data in the literature shows large disagreement in a few cases. The agreement with the theoretical predictions is also moderate for all model codes. The new versions of the TENDL library result mostly in better agreement with the experimental data. It is hard to explain the large disagreement of the earlier TENDL versions.  In case of ALICE and EMPIRE systematic overestimation can be observed.
Reliable experimental activation data are important for further improving the predictive power of the presently available theoretical codes. The present experimental data extend the energy range and the list reactions, for which comparison with the different model codes can be made, showing the need for better description of deuteron induced reactions in the nuclear model codes.

\begin{acknowledgements}
This work was done in the frame MTA-FWO research project and collaboration. The authors acknowledge the support of research projects and of their respective institutions in providing the materials and the facilities for this work. 
\end{acknowledgements}

\bibliographystyle{spphys}       
\bibliography{Pbd}   

%
%

\end{document}